\documentclass[prb, aps, floatfix, reprint, superscriptaddress, titlepage, amssymb, amsmath]{revtex4-1}

\usepackage{graphicx}
\usepackage{times}
\usepackage[lflt]{floatflt}
\def\be{\begin{equation}}
\def\ee{\end{equation}}
\def\bea{\begin{eqnarray}}
\def\eea{\end{eqnarray}}

\def\asi{{\it a}-Si}
\def\age{{\it a}-Ge}
\def\asih{{\it a}-Si:H}
\def\deg{$^\circ$}

\begin{document}
\title{
Effect of low-temperature annealing on the void-induced microstructure in amorphous silicon: A computational study
} 

\author{Durga Paudel}
\affiliation{Department of Physics and Astronomy, The University 
of Southern Mississippi, Hattiesburg, MS 39406}
\email[Corresponding author:\:]{Partha.Biswas@usm.edu}

\author{Raymond Atta-Fynn}
\affiliation{Department of Physics, University of Texas, Arlington, TX 76019}

\author{David A. Drabold}
\affiliation{Department of Physics and Astronomy, Ohio University, Athens, Ohio 45701}

\author{Parthapratim Biswas}
\affiliation{Department of Physics and Astronomy, The University of Southern Mississippi, Hattiesburg, MS 39406}

\begin{abstract}
We present a computational study of the void-induced microstructure in amorphous silicon ({\asi}) by generating ultra-large models of {\asi} 
with a void-volume fraction of 0.3\%, as observed in 
small-angle x-ray scattering (SAXS) 
experiments.  The relationship between the morphology of 
voids and the intensity of scattering in SAXS has been studied 
by computing the latter from the Fourier transform of the 
reduced pair-correlation function and the atomic-form factor 
of amorphous silicon.  
The effect of low-temperature ($\le$ 600 K) annealing on the scattering intensities 
and the microstructure of voids has been addressed, with 
particular emphasis on the shape and size of the voids, by studying 
atomic rearrangements on the void surfaces and computing the average 
radius of gyration of the voids from the spatial distribution of 
surface atoms and the intensity plots in the Guinier approximation. 
The study suggests that low-temperature annealing can lead to 
considerable restructuring of void surfaces, which is clearly 
visible from the three-dimensional shape of the voids but it may 
not necessarily reflect in one-dimensional scattering-intensity plots. 
\end{abstract}
\maketitle
 
\section{Introduction}

The microstructure of small- and large-scale inhomogeneities, 
such as multi-vacancies and voids, plays an important role in the determination 
of structural and optoelectronic properties of pure and hydrogenated 
amorphous silicon~\cite{street_1991}. The presence of voids in 
amorphous silicon ({\asi}) and germanium ({\age}) was postulated in 
the late 1960s by Brodsky 
and Title \cite{ESR1969} and Moss and Graczyk,\cite{Moss} with the aid 
of electron spin resonance (ESR) and small-angle electron diffraction 
measurements, respectively, to explain the low mass density of 
amorphous Si/Ge (by about 10--15\%) from their crystalline 
counterparts. In the intervening decades, a number of experimental techniques, 
ranging from small-angle x-ray scattering (SAXS),\cite{DQMahan}
spectroscopic ellipsometry (SE), Fourier transform infrared 
spectroscopy (FTIR),\cite{SEFTIR} and effusion of hydrogen 
and implanted-helium measurements \cite{Effusion2004} in {\asih} to 
nuclear magnetic resonance (NMR),\cite{NMR1985} provided 
an impressive database of experimental information on 
structural properties of {\asi} and its hydrogenated 
counterpart. By contrast, until recently,~\cite{PRB2018}
computational efforts \cite{PRA2017,SRE-Biswas,JAP2014,RBen} to address 
structural and electronic properties of {\asi} have been 
mostly limited to results obtained from small atomistic models, 
consisting of a few to several 
hundreds of atoms, depending upon the quantum-mechanical 
or classical nature of atomic interactions employed in building those 
models. Thus, the computational modeling of large-scale inhomogeneities in amorphous networks, such 
as voids in {\asi} and {\age}, were particularly hindered in the 
past due to the absence of large realistic models, which were needed to 
take into account the size and the number density of the voids, as 
observed in SAXS, NMR and FTIR studies. Since voids are 
nanoscale inhomogeneities, with a diameter ranging from 
10--40 {\AA} and a void-volume fraction of 0.1--0.4\%, it is 
necessary to produce ultra-large models, consisting of $10^5$ 
atoms or more, so that the models are not only experimentally 
compliant but also statistically reliable and capable of producing 
configurational-averaged values of physical properties. 
In this communication, we address the structure of void-induced morphology 
in {\asi} by carrying out realistic simulations 
of voids, using classical molecular-dynamics simulations. 

Of particular interest is the evolution of the microstructure 
of voids in {\asi}/{\asih}, induced by thermal annealing. While 
it is known that annealing at low temperature increases structural ordering in {\asi}, as 
observed in Raman spectroscopy,\cite{Raman} the effect of annealing 
on the shape and size of the voids is still largely unknown. Since the 
restructuring of atoms in the vicinity of void surfaces, especially 
in {\asih}, is intrinsically connected with the photovoltaic behavior 
of {\asih}-based solar devices, computational studies of the relaxation 
behavior of atoms on and near the void surfaces are of paramount importance. 
Here, we present a quantitative study of characterization of void 
shapes and sizes upon low-temperature annealing by examining the spatial 
distributions of void-surface atoms and computing the scattering 
intensity in SAXS, produced by the voids of evolving shape present 
in the models. 

The rest of the paper is as follows. In sec.\,II, we provide a brief 
description of the computational method to obtain atomistic models of 
{\asi} via classical molecular-dynamics (MD) simulations using the 
modified Stillinger-Weber potential.\cite{SWVink}  The computation of the scattering 
intensity in SAXS using the MD models as the structural basis is 
also described in this section.  Section III discusses the results 
of low temperature annealing, from 300 K to 600 K, on the 
void surfaces with an emphasis on the shape of the scattering 
intensity plots. The conclusions of our work are presented in sec.\,IV.

\section{Computational Method}
Since scattering contributions in SAXS chiefly originate from the 
small wave-vector region ($k \le 1$ {\AA}$^{-1}$), it is essential to extract information 
on the extended-scale inhomogeneities by generating large models of 
{\asi} for realistic simulation of scattering intensity in the small-angle 
region. Toward that end, we generated two independent {\asi} 
models (M1 and M2), consisting of 262 400 atoms in a cubic supercell of length 
176.123 {\AA}, using the modified Stillinger-Weber potential \cite{SWVink} 
via molecular-dynamics (MD) simulations in canonical ensembles. 
The MD simulations comprised the following steps: a) equilibration 
of an initial random configuration at 1800 K for 40 ps; b) gradual 
cooling of the equilibrated system from 1800 K to 300 K in steps 
of 100 K per 25 ps so that the total simulation time is 415 ps 
per cooling cycle; c) repetition of step (b) for up to 40 cycles until a 
high-quality amorphous network could form, which was characterized 
by a narrow bond-angle distribution and a small percentage of 
coordination defects;  d) total-energy relaxations of the system 
at 300 K, using the limited-memory BFGS method, \cite{BFGS1} to obtain the 
nearest local minimum-energy configuration.  

In order to study the temperature 
dependence of SAXS intensity and void shape, we generated four 
additional models from M1 and M2 by introducing ellipsoidal voids 
such that the total void-volume fraction, $f_v$, corresponds to 0.3\%, 
as observed in experiments.~\cite{William1989}
The ellipsoidal voids were generated randomly so
that they were sparsely distributed within amorphous 
networks with a minimum surface-to-surface distance of 6 {\AA} 
between any two neighboring voids. An axial ratio of 
$a:b:c = 0.5:1:2$, with a length scale of $b$ = 6 {\AA}, was 
employed to generate ellipsoidal voids and the silicon 
atoms in the region between $x$ and $x+d$, where $x = (a,b,c)$ 
and $d$ = 2.8 {\AA}, were marked as 
surface atoms in order to monitor the evolution of void 
shape upon annealing and relaxation. Two independent sets of 
models were studied to examine the effect of annealing on 
void shape and the intensity of scattering at 400 K and 
600 K. The corresponding relaxed models were also examined 
in our work for a comparison. 

The computation of scattering intensity in SAXS follows directly 
from the static structure factor, $S(k)$, and the atomic-form 
factor, $f(k)$, of {\asi}.  By invoking the homogeneous 
approximation, and subtracting the contribution 
coming from $k$ = 0, the structure factor can be 
written as,
\be
S(k)= 1 + \frac {1}{k}  \int _0^{R_{c}} G(r)\,\sin{kr}\, dr. 
\label{EQ2}
\ee
\noindent
Here, $G(r)$ is the reduced pair-correlation function (PCF) and $k$ is 
the magnitude of wave-vector transfer, $k = 4\pi\sin(\theta)/\lambda$, 
where $2\theta$ is the scattering angle and $\lambda$ is the 
wavelength of x-ray. The upper limit 
of the integral in Eq.\,(\ref{EQ2}) is chosen so that $G(r)$ 
can be neglected for $r > R_c$ = 30 {\AA}. Denoting $f(k)$ as 
the atomic-form factor, the scattering intensity, $I(k)$, 
can be written as, $I(k) = \frac{1}{N} |f(k)|^2 S(k)$, where $N$ 
is the total number of atoms. 

\section{Results and Discussions} 
\begin{table}
\centering
\caption{
{\small 
Structural properties of simulated models of amorphous silicon: $N$, 
$\rho$, $C_{4}$, $C_{3}$, $r_{\mathrm{avg}}$, 
$\theta_{\mathrm{avg}}$, and $\Delta\theta$ are the size of the 
simulation cell, total number of atoms, density, four-fold 
coordination, three-fold coordination, average bond length, 
average bond angle, and the RMS deviation of the bond angles, 
respectively.
}
\label{TAB1}
}
\begin{tabular}{@{}l*{8}{c}} 
\hline
Model & $N$ & $\rho$ (g/cm$^{3}$) & $C_{4}$ (\%) &$C_{3}$ (\%) &  
$r_{\mathrm{avg}}$ ({\AA}) &  $\theta_{\mathrm{avg}}$ &  $\Delta\theta$ \\
\hline 
\hline
M1 & 262 400 & 2.24 &  97.65 & 1.27 & 2.39  & 109.24{\deg} &  9.10{\deg}     \\ 
M2 & 262 400 & 2.24 &  97.71 & 1.27 & 2.39  &  109.25{\deg} &  9.08{\deg}      \\   
\hline
\end{tabular}
\end{table}

\begin{figure}[ht!]
\includegraphics[width=0.45\textwidth]{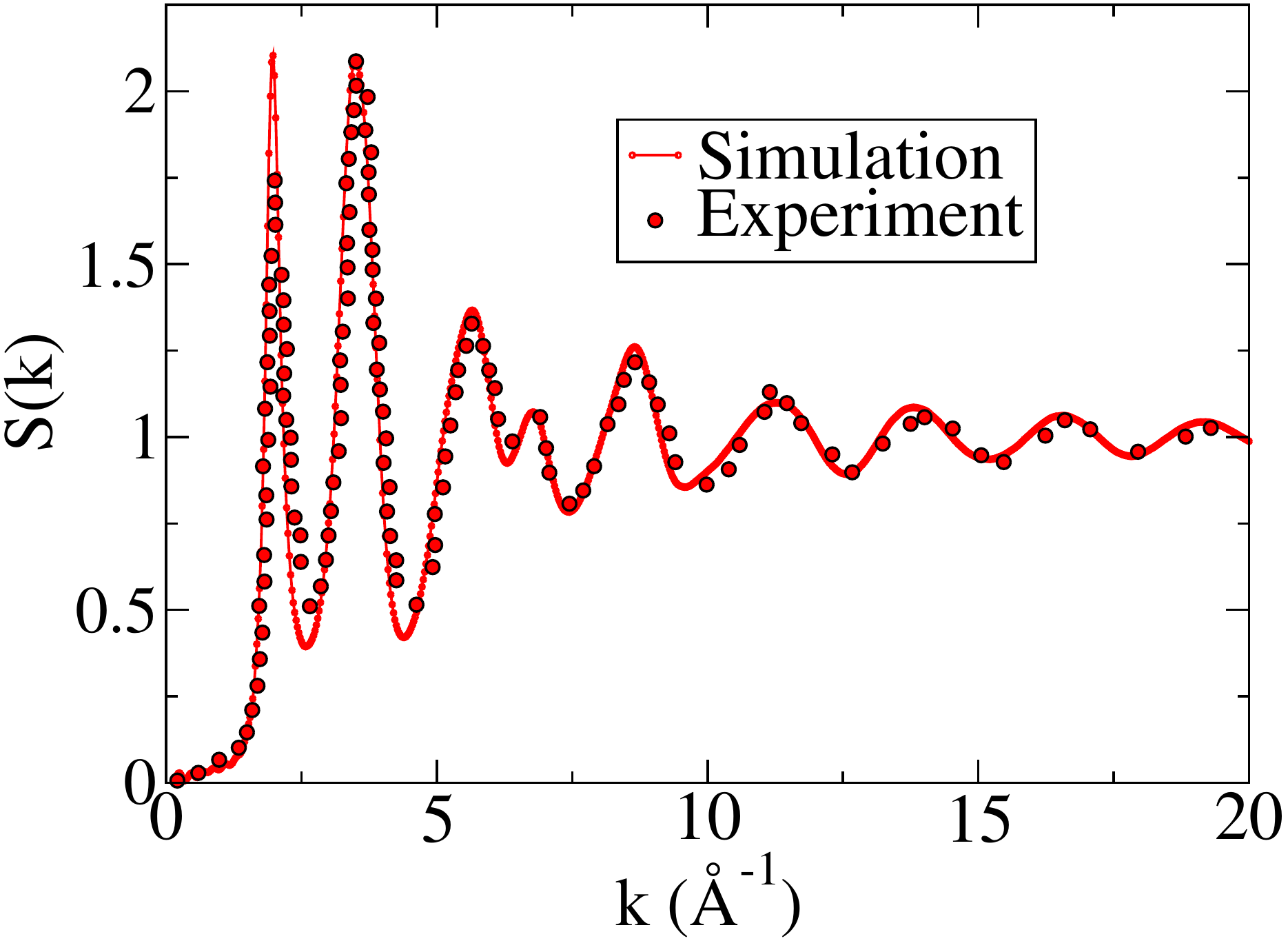}
\caption{
{\small 
The static structure factor of {\asi} from simulations and 
experiments.  The simulation data were averaged over M1 
and M2 models.  Experimental data correspond to annealed 
samples of {\asi} from Ref.\,\onlinecite{Laaziri}. 
}
\label{FIG1}
}
\end{figure}

Table \ref{TAB1} lists the characteristic structural properties of M1 
and M2 models, obtained from molecular-dynamics (MD) simulations. 
Given the ultra-large size of the models, it is remarkable that 
the concentration of coordination defects and the root-mean-square (RMS)
deviation of bond angles are less than 2.4\% and 
10{\deg}, respectively. Figure \ref{FIG1} shows the static structure 
factor of the simulated models, averaged over M1 and M2 models, 
along with the experimental structure-factor data from wide-angle 
x-ray diffraction measurements.~\cite{Laaziri} The simulated 
values of the structure factor match closely with the experimental 
data, indicating that the models are not only fully consistent 
with two- and three-body correlation functions but also have a low 
concentration of coordination defects. To our knowledge, these ultra-large 
models are among the very best of {\asi} models, from classical MD simulations,  
with 10$^5$ or more atoms so far reported in the literature.

\begin{figure}[t!]
\includegraphics[width=0.3\textwidth]{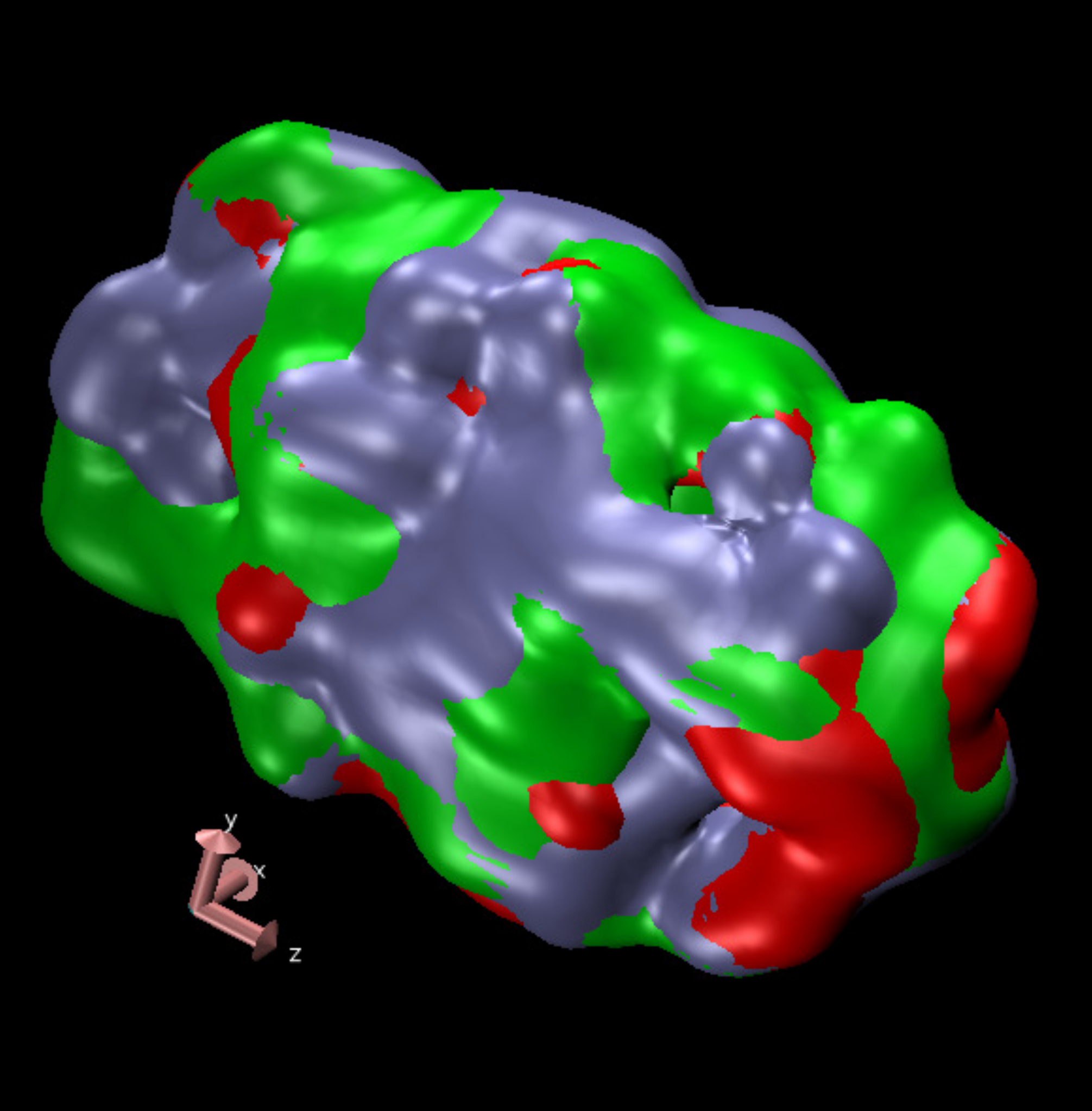}
\caption{
\label{FIG2}
{\small 
The changing shape of an ellipsoidal void (M1-V5) in M1 model 
at two different annealing temperatures.  Red and green 
colors correspond to the shape at temperature 600 K and 
400 K, respectively, along with the original ellipsoid 
surface shown in blue color. The color patches are indicative 
of surface reconstruction, showing the evolving shape of 
the surface with temperature. The symbol M1-V5 indicates 
void number 5 in M1 model. 
}
}
\end{figure}

\begin{figure}[t!]
\includegraphics[width=0.4\textwidth]{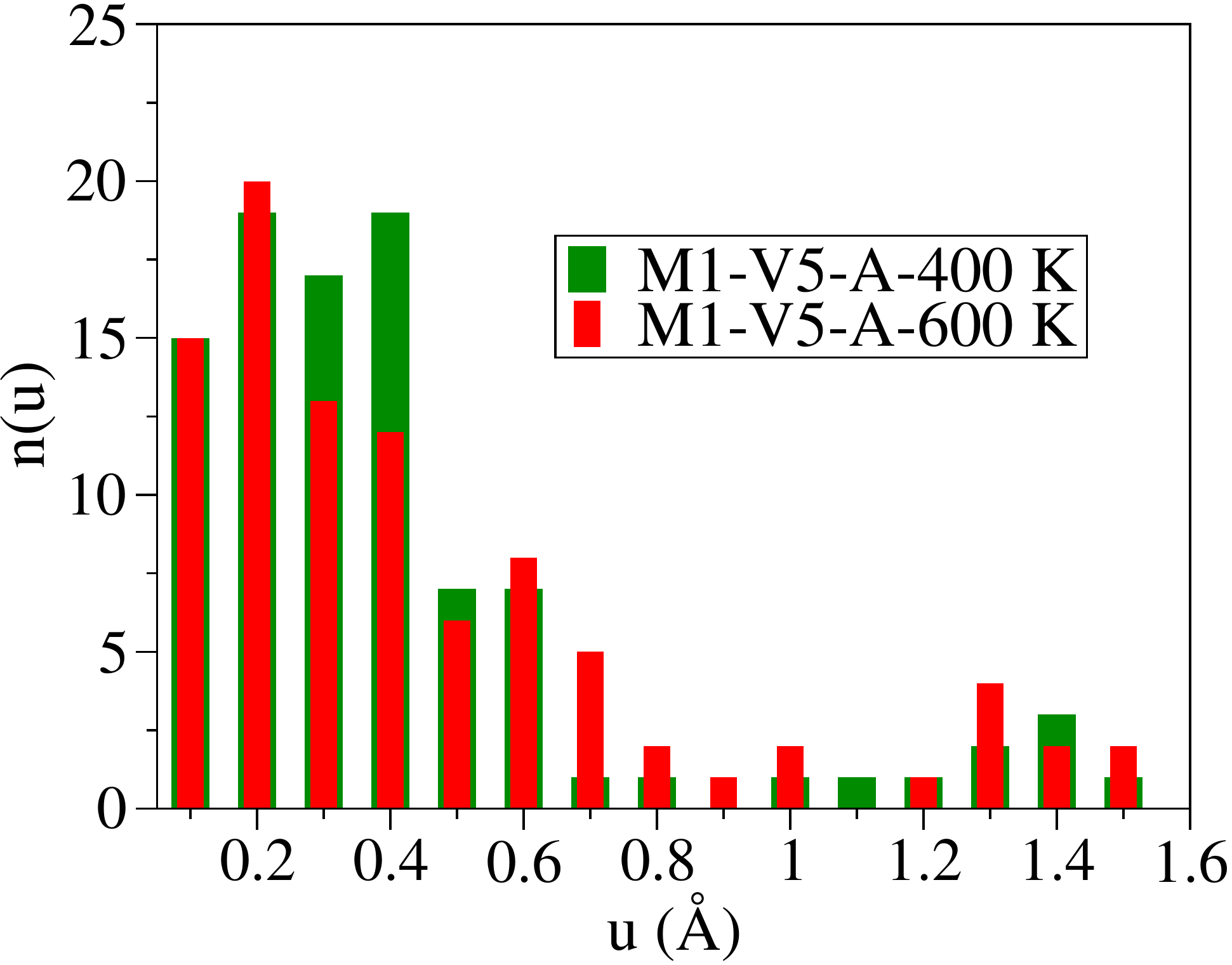}
\caption{
\label{FIG3}
{\small 
The distributions of atomic displacements, $n(u)$, of 
the surface atoms on the ellipsoidal void (M1-V5) shown 
in Fig.\,\ref{FIG2}. The distributions at 600 K and 
400 K are shown in red and green colors, 
respectively. 
} 
}
\end{figure}

The effect of low-temperature annealing on the shape and size of the 
voids was studied by analyzing the results from real and 
reciprocal spaces. Figure \ref{FIG2} depicts the evolution 
of an ellipsoidal void at 400 K and 600 K, along with 
the original shape, all superimposed on each other. The 
distorted ellipsoidal surfaces observed at 400 K and 600 K 
are shown in green and red colors, respectively, whereas 
the original void surface is shown in blue color. 
The striking color patches, which originate from the superposition 
of three ellipsoidal surfaces, are reflective of the degree of 
reconstruction of the surface at 400 K (green) and 600 K (red). These surfaces 
were generated from the position coordinates of the respective 
void-surface atoms, using an identical set of iso-surface parameters in 
{\sc VMD} software.~\cite{VMD}
To obtain a quantitative estimate of the degree of surface reconstruction, the 
distributions of atomic displacements ($u$) of the surface atoms 
of the ellipsoidal void at 600 K and 400 K are plotted in Fig.\,\ref{FIG3}. 
While it is evident from Fig.\,\ref{FIG3} that majority of the 
surface atoms have displaced from their original position by a 
distance of 0--0.75 {\AA}, a small number of atoms have moved more 
than 1.0 {\AA}, causing considerable restructuring of the void surface. 
This observation has been found to be true for other ellipsoidal voids 
as well. 

\begin{figure}[ht]
\includegraphics[width=0.4\textwidth]{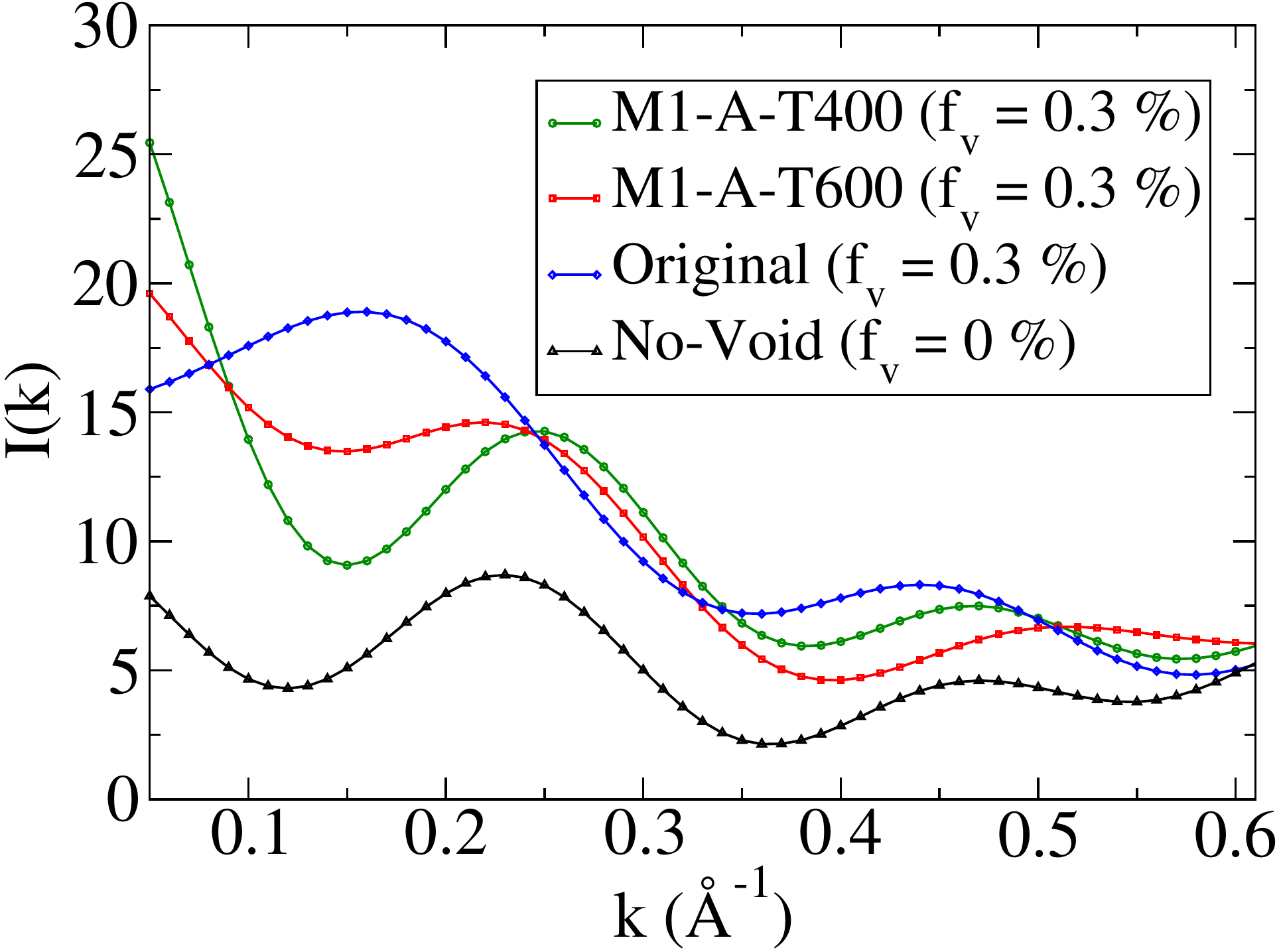}
\caption{
\label{FIG4}
{\small 
The computed values of the scattering intensity from the 
original model (blue) before annealing and after annealing 
at 400 K (green) and and 600 K (red).  For comparison, the 
scattering intensity from the model with no voids (black) 
is also shown in the plot. 
}
}
\end{figure}

While an analysis of the three-dimensional distribution of atoms clearly 
reveals atomic rearrangements on void surfaces, a question of great 
importance, from the point of view of SAXS, is to what 
extent these temperature-induced structural changes can be reflected 
in $I(k)$-vs-$k$ plots. 
Since SAXS essentially provides {\it scalar} information or {\it one-dimensional} 
physical quantities, such as scattering intensities and the average 
radius of gyration of the voids in the Guinier approximation,~\cite{Guinier-book} 
it is pertinent to examine whether these scalar quantities are sensitive 
to structural changes observed on the void surfaces. Figure \ref{FIG4}
shows the scattering intensities 
obtained from the original model with voids (blue) and the corresponding 
annealed models at 600 K (red) and 400 K (green). For comparison, we 
have also included the intensity values obtained from the original 
model with voids (blue) and without voids (black) before annealing.  
A few observations immediately follow from Fig.\,\ref{FIG4}. 
First, an excess of small-angle scattering is apparent in the computed values 
of the scattering intensity from the models with 
voids (red/green/blue), compared with the one with 
no voids (black). Second, the variation of the scattering 
intensity in the presence of voids is mostly pronounced in the 
small-$k$ region below 0.25 {\AA}$^{-1}$. Since the wave-vector 
region below 0.1 {\AA}$^{-1}$ is not very reliable, owing to 
the presence of numerical noise at large distances in $G(r)$ 
(see Eq.\,\ref{EQ2}), it would not be inappropriate to conclude 
that the changes are somewhat weakly reflected in the intensity 
plots as far as our simulations are concerned. Third, it 
appears that the scattering plot at 600 K (red) is somewhat 
smoothed out compared to the one at 400 K (green), which can be 
attributed to the temperature-induced improved local ordering 
in the network, especially in the vicinity of the void regions. 
Finally, the minor changes in the shape of the intensity plot are 
consistent with the values of the radius of gyration, obtained 
from the {\it real-space} distribution of surface atoms.  

\begin{figure}[ht!]
\includegraphics[width=0.35\textwidth]{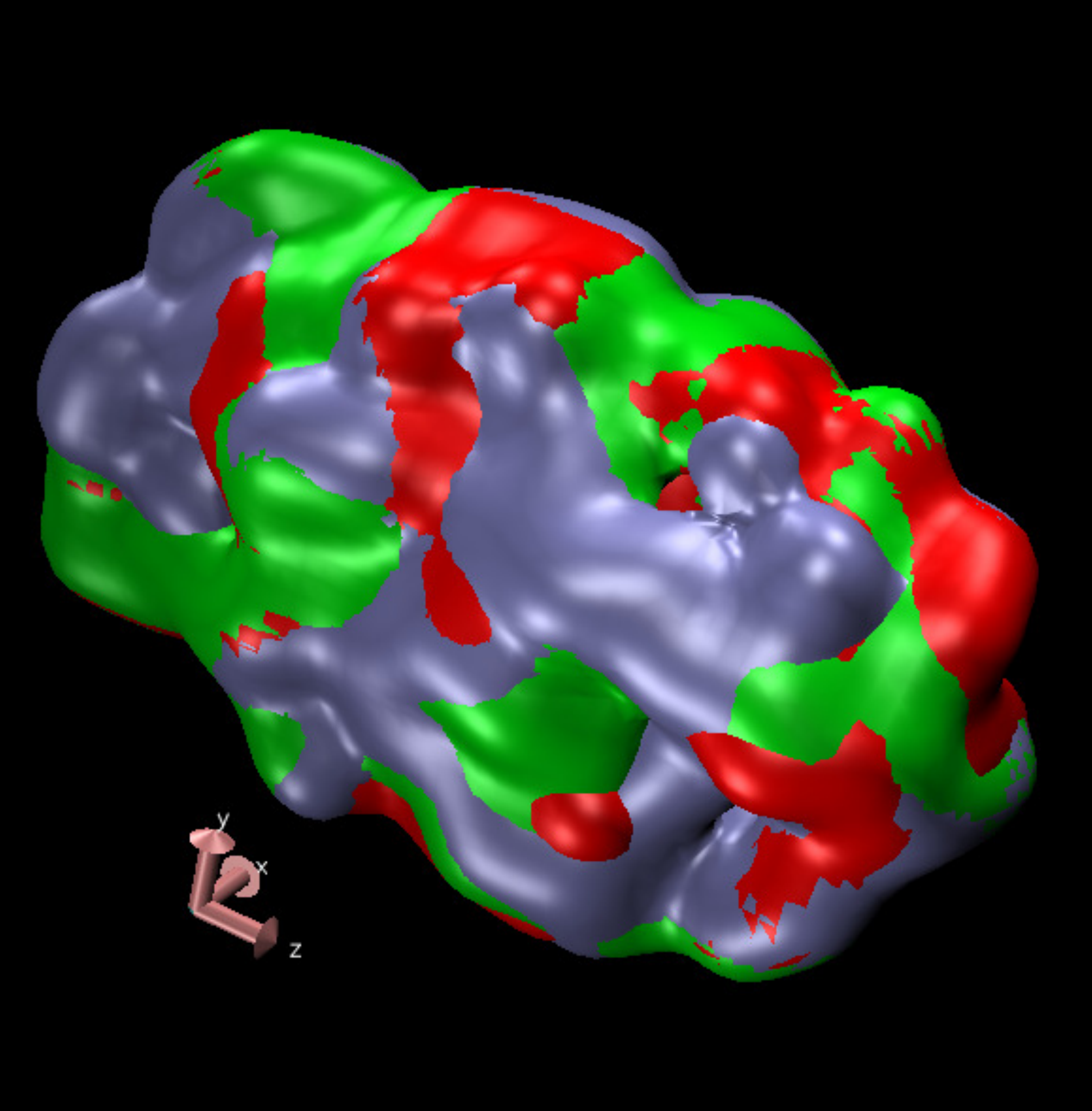}
\caption{\label{FIG5}
{\small 
The shape of the same ellipsoidal void obtained after relaxation of 
the annealed model (shown in Fig.\,\ref{FIG2}) at 400 K (green) 
and 600 K (red). 
}
}
\end{figure}

\begin{figure}[t!]
\includegraphics[width=0.35\textwidth]{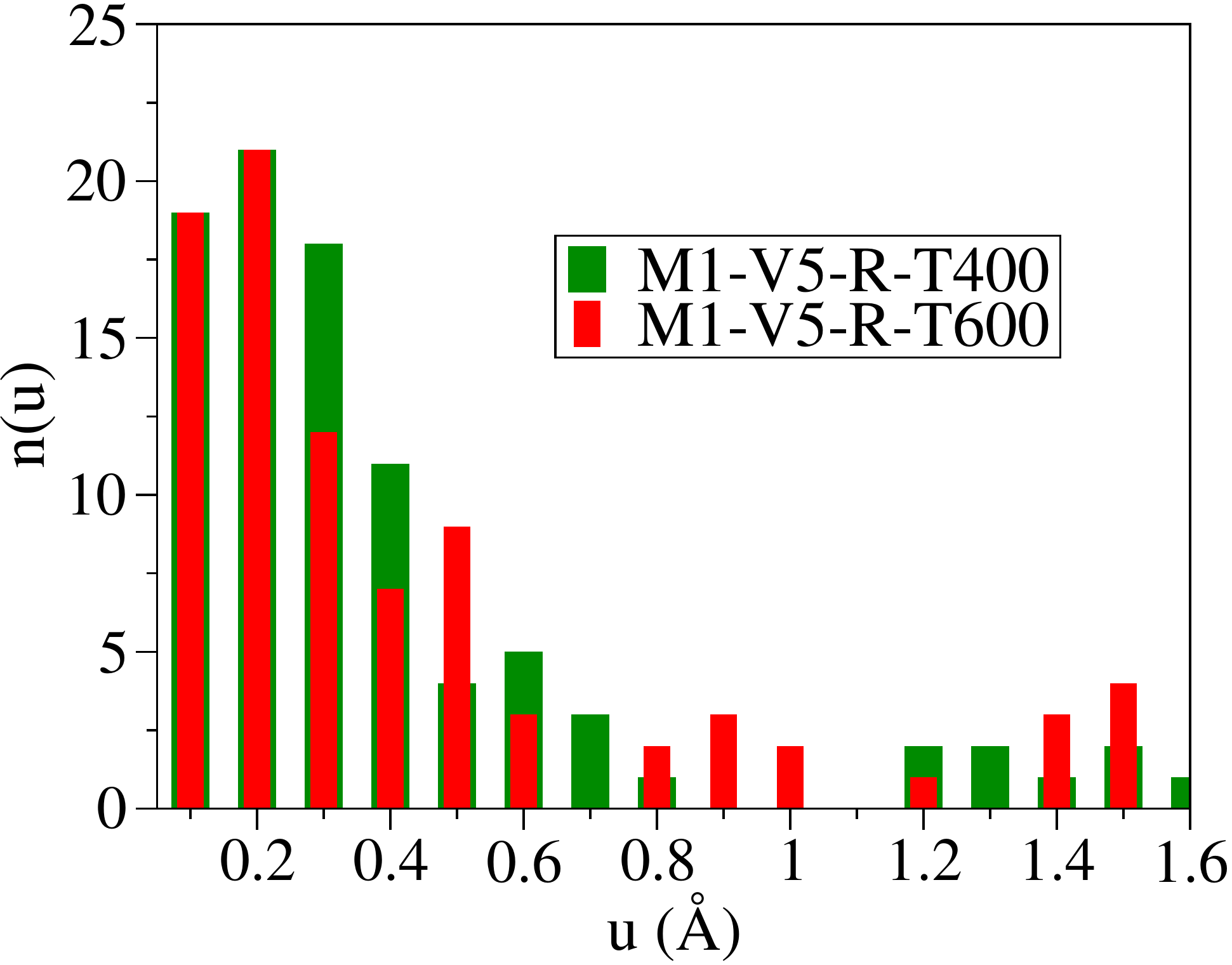}
\caption{\label{FIG6}
{\small 
The distributions of atomic displacements ($u$) of 
the atoms on the ellipsoidal void surface with the same 
color code as in Fig.\,\ref{FIG5}. The symbol M1-V5-R-T400 
signifies the relaxation of void number 5 in M1 model 
after annealing at 400 K. 
}
}
\end{figure}

\begin{figure}[ht!]
\includegraphics[width=0.40\textwidth]{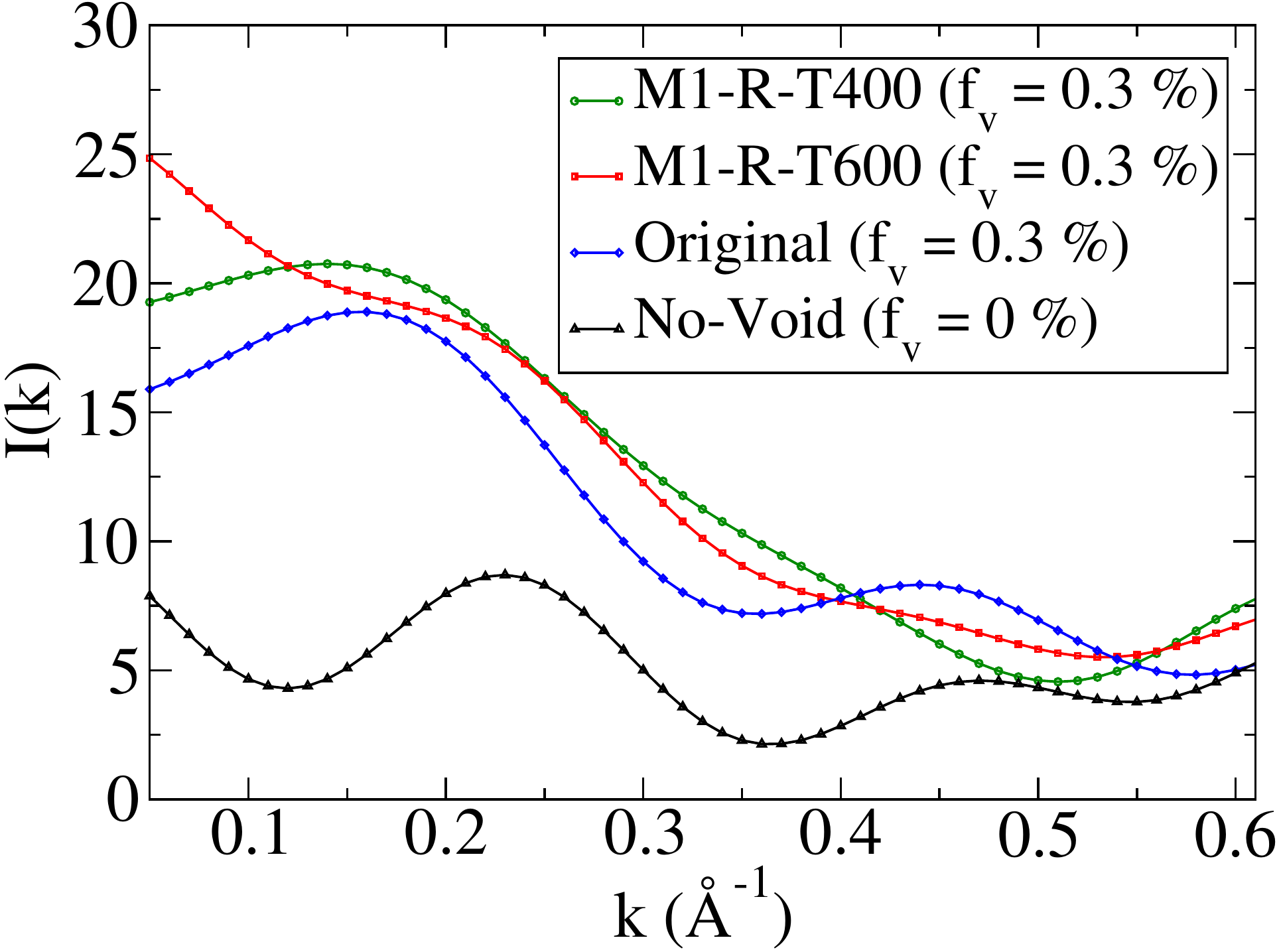}
\caption{\label{FIG7}
{\small 
The scattering intensities from the relaxed models after annealing 
at 400 K (green) and 600 K (red). The intensity values obtained 
from the original model with no voids (black) are also plotted for comparison. 
}
}
\end{figure}

Table \ref{TAB2} lists $R_g$ values obtained from the spatial 
distribution of void-surface atoms at 400 K and 600 K, which 
are quite close to each other. On the other hand, the radius 
of gyration, $r_g$, obtained from the intensity plots in the 
Guinier approximation,~\cite{Guinier-book} shows a noticeable 
variation with temperature.  
Since the Guinier approximation entails expressing 
$I(k) \approx I(0)\,\exp(-k^2\,r_g^2/3)$ in the limit $k\to 0$, and 
fitting the intensity values on a semi-log plot, $\ln I(k)$-vs-$k^2$, 
$r_g$ values obtained from 
this approach are quite sensitive to the computed values 
of the intensity and the length of the small-$k$ region employed 
in the calculations. Here, we have employed intensity values in 
the region 0.015 $\le k^{2}\le$ 0.16 {\AA}$^{-1}$, to obtain 
the $r_g$ values listed in Table \ref{TAB2}. In Figs.\,\ref{FIG5} 
to \ref{FIG7}, we have presented the corresponding shape, atomic 
displacements, and the values of the scattering intensities 
obtained by relaxing the annealed models at temperature 400 K 
and 600 K, respectively.  Although considerable changes in the shape of the 
relaxed ellipsoidal void can be seen in Fig.\,\ref{FIG5}, the 
scattering intensities in Fig.\,\ref{FIG7} appear to be not 
particularly susceptible to the changes observed on the void 
surfaces.  \\

\begin{table}[ht] 
\centering
\caption{
{\small 
Characteristic properties of void distributions in annealed models of 
{\asi}. $R_{g}$, $\sigma$, and $r_{g}$ indicate the average 
gyrational radius, the root-mean-square fluctuation of $R_g$, 
and the Guinier radius from the scattering intensity plots in 
Fig.\,\ref{FIG4}, respectively. A/R-T400 corresponds to the 
annealed/relaxed model at 400 K.
\label{TAB2} 
}
} 
\begin{tabular}{@{}l*{10}{l}}
\hline
Models & $R_{g}\pm\sigma$ ({\AA}) & $r_{g}$ ({\AA}) &Models & $R_{g}\pm\sigma$ ({\AA}) 
& $r_{g}$ ({\AA})\\
\hline 
M1-A-T400 & 5.87$\pm$0.12 & 3.64 & M1-R-T400 & 5.88$\pm$0.11 & 4.42\\  
M1-A-T600 & 5.84$\pm$0.12 & 5.19 & M1-R-T600 & 5.86$\pm$0.12 & 4.85\\ 
M2-A-T400 & 5.94$\pm$0.07 & 4.66 & M2-R-T400 & 5.96$\pm$0.08 & 4.53\\ 
M2-A-T600 & 5.92$\pm$0.1  & 5.87 & M2-R-T600 & 5.94$\pm$0.1  & 5.33\\ 
\hline
\end{tabular}
\end{table}

\section{Conclusions}
In this paper, we have studied the microstructure of nanometer-size 
voids, by simulating ultra-large models of {\asi}, using 
classical molecular-dynamics simulations.  The intensity of 
small-angle x-ray scattering, produced by ellipsoidal voids 
present in the models, is computed from the Fourier transform 
of the reduced pair-correlation function and the atomic-form 
factor of silicon in an effort to study the effect of low-temperature 
annealing of up to 600 K on the morphology, especially the shape 
and size, of the voids.  Our study shows that thermalization of 
the models at 600 K produces considerable restructuring of void 
surfaces with atomic displacements of the surface atoms as high 
as 1.6 {\AA}. The reconstruction of void surfaces has been found 
to be somewhat weakly reflected in one-dimensional scattering plots, obtained 
from the simulated values of the scattering intensity produced 
by the models. It has been observed that, despite considerable 
changes of void shape, one-dimensional measures of size, such 
as gyrational and Guinier radii -- obtained from the spatial 
distribution of atoms and the intensity plots, respectively -- 
are not particularly affected by low-temperature annealing. 

\section*{Acknowledgements}
This work is partially supported by U.S. National Science 
Foundation (NSF) under Grants No. DMR 1507166, No. DMR 1507118, 
and No. DMR 1506836.

\section*{References}
%

\end{document}